%% file: spatially_adaptive_phasefield_model.tex
\documentclass[review]{elsarticle}
\include{package}
\usepackage{hyperref}

\biboptions{authoryear}

\begin{document}

\begin{frontmatter}

\title{A spatially adaptive phase-field model of fracture}


\author[address1,address2]{Dhananjay Phansalkar \corref{mycorrespondingauthor}}
\cortext[mycorrespondingauthor]{Corresponding author}
\ead{dhananjay.phansalkar@fau.de}
\author[address3]{Kerstin Weinberg}
\author[address4]{Michael Ortiz}
\author[address1]{Sigrid Leyendecker}

\address[address1]{Institute of Applied Dynamics, Friedrich-Alexander-Universit\"at Erlangen-Nürnberg, Immerwahrstrasse 1, Erlangen 91058, Germany}
\address[address2]{Central Institute for Scientific Computing, Friedrich-Alexander-Universit\"at Erlangen-Nürnberg, Martensstrasse 5a, Erlangen 91058, Germany}
\address[address3]{Chair of Solid Mechanics, Universit\"at Siegen, Paul-Bonatz-Strasse 9-11, Siegen 57076, Germany }
\address[address4]{Division of Engineering and Applied Sciences, California Institute of Technology, Pasadena, CA 91125, USA}

\begin{abstract}
Phase-field models of fracture introduce smeared cracks of width commensurate with a regularisation length parameter $\epsilon$ and obeying a minimum energy principle. Mesh adaptivity naturally suggests itself as a means of supplying spatial resolution were needed while simultaneously keeping the computational size of the model as small as possible. Here, a variational-based spatial adaptivity is proposed for a phase-field model of fracture.

The conventional phase-field model is generalised by allowing a spatial variation of the regularisation length $\epsilon$ in the energy functional. The optimal spatial variation of the regularisation length then follows by energy minimisation in the same manner as the displacement and phase fields. The extended phase-field model is utilised as a basis for an adaptive mesh refinement strategy, whereby the mesh size is required to resolve the optimal length parameter locally. The resulting solution procedure is implemented in the framework of the finite element library FEniCS. Selected numerical tests suggest that the spatially adaptive phase-field model exhibits the same convergence rate as the conventional phase-field model, albeit with a vastly superior constant, which results in considerable computational savings.
\end{abstract}


\end{frontmatter}


\section{Introduction}
Understanding a material's fracture is essential to design mechanical systems, particularly for components with a low safety factor.
However, reliable fracture computations are challenging for numerous reasons, like the tracking of complex crack paths, the singularity of stress at the crack tip, and the dependency of the solution on the discretisation method and the mesh size. To overcome these difficulties, non-local methods like phase-field models have gained popularity. In such a method, the crack is 'smeared' over a certain width $\epsilon$, regularising the sharp interface problem of the discrete crack. Phase-field models are easy to use and implement, and, in contrast to discrete crack models, they do not require mesh manipulation for crack initiation, kinking, and branching.

Phase-field models are typically derived from a variational principle \cite{bourdinVariationalApproachFracture08a}. The direct, monolithic solution of the resulting fully-coupled non-linear Euler-Lagrange equations has problems usually solved with a viscous regularisation \cite{miehePhaseFieldModel10}, an optimal damping in the Newton update \cite{gerasimovLineSearchAssisted16a}, inertial terms \cite{bordenPhasefieldDescriptionDynamic12}, or simply by using a  staggered approach \cite{miehePhaseFieldModel10,Bilgen_etal2018,hirshikeshPhaseFieldModelling19}. Furthermore, for accurate fracture computations, the phase-field method must be supplemented by further modelling details, for example, to guarantee the irreversibility of crack growth \cite{kuhnContinuumPhaseField10,miehePhaseFieldModel10}, to account for the loss of stiffness of the damaged material \cite{sargado2018high},
and to derive the tension-related crack-driving force from the energy form \cite{ambatiReviewPhasefieldModels15,BilgenWeinberg2019_CMAME}. In this work, we focus on the length parameter $\epsilon$ determining the non-local crack regularisation solely.

It is shown that in the limit for the length parameter $\epsilon \to 0$ the phase-field energy functional approaches the discrete brittle fracture energy in the sense of $\Gamma$-convergence \cite{giacominiAmbrosioTortorelliApproximationQuasistatic05}. This implies choosing a relatively small $\epsilon$ for a good approximation of the discrete solution. Various studies \cite{linseConvergenceStudyPhasefield17,bourdinVariationalApproachFracture08a} suggest that a relatively fine mesh size relative to $\epsilon$ is needed, i.\,e., $h<\epsilon$. Thus, without any adaptive refinement for a complex crack path or without  \textit{prior} knowledge of the crack path, a uniform fine mesh is required leading to a high computational cost. Additionally, in the recent work \cite{pandolfiComparativeAccuracyConvergence21}, a numerical study shows there exist an optimal $\epsilon$ for a given mesh size $h$, in particular $h \propto \epsilon^2$. Hence, it is challenging and, at the same time, crucial to choose the appropriate regularisation length parameter $\epsilon$ and mesh size $h$. An inappropriate combination of $\epsilon$ and $h$ may lead to difficulty reproducing results and non-optimal solutions for a given pair of $\epsilon$ and $h$.

This work strives to resolve these challenges by introducing a variational formulation to determine an optimal field $\epsilon(\bx)$ and extending it with a corresponding mesh refinement strategy. In Section \ref{sec: phasefield model}, a modified energy functional is introduced to compute the optimal regularisation length parameter.
We formulate two variants; the first yields an optimal uniform length variable $\epsilon$ for the complete domain, while in the second variant, $\epsilon(\bx)$ is considered as a spatially varying length variable which is determined together with the displacement and phase-field. Numerical computations are provided in Section \ref{sec:fem}.
The spatially varying length variable $\epsilon(\bx)$ is extended with a corresponding mesh refinement algorithm in Section \ref{sec:refinement method}. Section \ref{sec:numerical implementation} briefly explains the numerical implementation along with various parameters used in the study, and in Section \ref{sec:results} a convergence study is presented. Concluding remarks and an outlook close the paper in Section \ref{sec:summary}.

\section{Modified phase-field formulation of fracture} \label{sec: phasefield model}
In brittle fracture, the total energy of the mechanical system is composed of the elastic strain energy and a fracture energy contribution. In phase-field modelling, this total energy includes local dissipation and material degradation functions. Although different possible formulations exist, the conventional model assumes quadratic dissipation and degradation
\cite{ambrosioApproximationFunctionalDepending90,bourdinVariationalApproachFracture08a,miehePhaseFieldModel10,Negri2013}.
In this work, we follow this approach and consider the following energy functional:
\begin{align}
	E ( \bu (\bx) ,c(\bx)) &= \int_{\Omega}  \frac{1}{2} [1-c(\bx)]^2  \psi(\bvarepsilon (\bu(\bx))) + \mathcal{G}_c \left[ \frac{c(\bx)^2}{2\epsilon}+\frac{\epsilon}{2} \vert \nabla c(\bx) \vert^2\right] \td\bx \label{eq: std functional} \\
	\psi(\bvarepsilon(\bu(\bx))) &= \frac{1}{2}\boldsymbol{\varepsilon}(\boldsymbol{u}(\bx)):[\mathbb{C}:\boldsymbol{\varepsilon}(\boldsymbol{u}(\bx))]
	=\frac{1}{2} \lambda [\text{tr}(\boldsymbol{\varepsilon}(\bu(\bx)))]^2 + \mu [\boldsymbol{\varepsilon}(\bu(\bx)):\boldsymbol{\varepsilon}(\bu(\bx))]
\end{align}
where $\boldsymbol{u}(\bx) $ is the displacement  and $c(\bx)$  the phase-field; $\boldsymbol{\varepsilon} = \operatorname{sym}(\nabla \bu(\bx)))$ represents the strain, $\psi$  isthe strain energy density, and $\epsilon$ is the regularisation length variable. For readability, we usually omit the arguments of the fields. The material is described by the elastic tensor $\mathbb{C}$ with Lam\'{e} parameters $\lambda$ and $\mu$ and the fracture toughness $\mathcal{G}_c$.

This work alleviates the issue of choice of $\epsilon $ by minimising the functional \eqref{eq: std functional} not just with respect to $\boldsymbol{u}$ and $c$ but also with respect to $\epsilon$ resulting in an optimal transition zone. However, to ensure the boundness and regularity of $\epsilon$ if the value of phase-field $c$ approaches zero, we introduce two terms in the functional, weighted with positive parameters $\eta$ and $\beta$. The resulting modified energy functional is now formulated as:
\begin{align}\label{eq:modified_energy_functional}
	E ( \boldsymbol{u} ,c, \epsilon) &= \int_{\Omega}  [1-c]^2  \psi(\boldsymbol{\varepsilon}) + \mathcal{G}_c \left[ \frac{c^2 + {\eta}}{2\epsilon}+\frac{\epsilon}{2} \vert \nabla c \vert^2\right]  + {\beta} \epsilon \td\boldsymbol{x}
\end{align}
where $\epsilon$ now represents a regularisation length variable, $\beta$ and $\eta$ are  penalty and model parameter, respectively. The choice of $\eta$ and $\beta$ needs to ensure that $\epsilon$ is small at the crack-tip or in the regions of high stress intensity factor and large everywhere else. Nonetheless, this choice is not straightforward and will be discussed in Section \ref{section:parameter estimation}. There are two possible ways to minimise the functional \eqref{eq: std functional} with respect $\epsilon$, i.\,e., either with respect to a uniform $\epsilon$ for the complete domain or with respect to a spatially varying length variable $\epsilon(\bx)$. We explore these two cases in the following section.

\subsection{Optimal uniform length variable $\epsilon$}
We consider the first case of uniform optimal $\epsilon$ in the domain $\Omega$.
We minimise the functional \eqref{eq:modified_energy_functional} with respect to $\boldsymbol{u}(\bx)$, $c(\bx)$ and also with respect to $\epsilon$. This leads to the following Euler-Lagrange equations:
\begin{align}\label{eq:optimalELE1}
	\nabla \cdot\boldsymbol{\sigma} = \boldsymbol{0} \quad  &\text{in} \quad \Omega \\
	\frac{c \mathcal{G}_c}{\epsilon} -2 \left[ 1-c\right] \psi(\boldsymbol{\varepsilon}) -  \mathcal{G}_c \epsilon \Delta  c = 0 \quad &\text{in} \quad \Omega\\\label{eq:optimalELE3}
	\epsilon = \sqrt{\frac{\int_{\Omega}(c^2 +\eta)  \td\boldsymbol{x}}{\int_{\Omega}(|\nabla c|^2+\frac{2 \beta}{\mathcal{G}_c}) \td\boldsymbol{x}}} \quad &\text{in} \quad \Omega
\end{align}
where $\boldsymbol{\sigma} = [1-c]^2\mathbb{C}:\boldsymbol{\varepsilon}$ is the stress tensor. In addition to the two equations of the conventional phase-field model, the third equation provides an optimal uniform length variable $\epsilon$ in $\Omega$. Furthermore, we have the Dirichlet boundary conditions for displacement $\boldsymbol{u} =  \boldsymbol{u_0} \ \text{on} \ \partial \Omega_{\text{d}}$ and phase-field $c =  c_0 \ \text{on} \ \partial \Omega_{\text{p}}$. These boundary conditions correspond to a  load and a pre-existing crack, respectively. We also apply Neumann boundary conditions for the phase-field, $\nabla c \cdot \boldsymbol{n} = 0 \ \text{on} \ \partial \Omega \setminus \partial \Omega_{\text{p}}$, to smoothen the phase-field towards the boundary.

\subsection{Spatially-varying length variable $\epsilon(\bx)$}
The above model can  further be improved by considering $\epsilon$ as a field variable.
Thus minimizing the functional \eqref{eq:modified_energy_functional} with respect to $\boldsymbol{u}(\bx), c(\bx) $ and $\epsilon(\bx)$ results in the following Euler-Lagrange equations:
\begin{align}
	\nabla \cdot\boldsymbol{\sigma} = \boldsymbol{0} \quad  &\text{in} \quad \Omega \label{eq:stress} \\
	\frac{c \mathcal{G}_c}{\epsilon} -2 \left[ 1-c\right] \psi(\boldsymbol{\varepsilon}) -  \mathcal{G}_c \nabla \cdot\left[\epsilon \nabla  c\right] = 0 \quad &\text{in} \quad \Omega  \label{eq:phasefield} \\
	\epsilon = \sqrt{\frac{c^2 + {\eta}}{|\nabla c|^2 + \frac{2 {\beta}}{\mathcal{G}_c}}} \quad &\text{in} \quad \Omega \label{eq:spatial epsilon}
\end{align}
whereby the first two equations are similar to the standard phase-field model. Additionally, we have the third equation for the spatially varying length variable $\epsilon(\bx)$, which can be expressed explicitly in terms of $c$ and other numerical parameters.
These equations are then solved in combination with the boundary conditions introduced above.

\section{Finite-element implementation}\label{sec:fem}
One of the substantial benefits of phase-field models of fracture is a relatively simple numerical implementation. As such, the weak forms of the Euler-Lagrange equations
(\ref{eq:optimalELE1}-\ref{eq:optimalELE3}) and (\ref{eq:stress}-\ref{eq:spatial epsilon}) are  implemented in the finite element framework of FEniCS \cite{alnaesFEniCSProjectVersion15}.
Our specific implementation is based on the work of \cite{hirshikeshPhaseFieldModelling19}.
We use triangular elements with linear and quadratic shape functions for the displacement field and the phase-field, respectively, and first-order discontinuous elements for the spatially varying length variable $\epsilon(\bx)$.

For meaningful computations, it is required to resolve the regularisation length by the mesh. However, a uniformly fine mesh leads to high computational costs; hence, to have a proper balance between accuracy and cost, we need a criterion for local mesh refinement. This criterion can be a local error estimator based on configurational forces or other refinement indicators, see, e.g. \cite{mieheRobustAlgorithmConfigurationalforcedriven07,welschingerConfigurationalForceBasedAdaptiveFE10}. However, from the perspective of a phase-field model, it is reasonably intuitive to have the mesh size $h$ correlated to the regularisation length $\epsilon$. Thus, we use the spatially varying length variable $\epsilon(\bx)$ to develop an adaptive mesh refinement strategy.

\subsection{Adaptive mesh refinement strategy}\label{sec:refinement method}
In regions with a small length variable $\epsilon(\bx)$, the mesh is refined to achieve a mesh size that is smaller than a specific function of $\epsilon$, i.\,e., $h\leq f(\epsilon)$. The details of this function $f(\epsilon)$ are discussed in Section \ref{sec:fun of epsilon}. The refinement is performed in a loop over all triangular elements. For each element, we find the minimum value of $\epsilon(\bx)$, i.\,e., $\epsilon_e$. To decide whether a refinement of an element is required, the following criteria are examined. First, we check whether the value of $\epsilon_e$ is smaller than a specific threshold value $\epsilon_\text{refine}$. If this is fulfilled, we check whether the mesh size of an element $h_e$ is greater than $f(\epsilon_e)$. Additionally, we constrain the minimal mesh size, i.\,e., $h_e$ is bounded from below with $h_\text{min}$. This process is summarised in Algorithm~\ref{alg: mesh refinement}.

\begin{algorithm}[H]
	\label{alg: mesh refinement}
	\SetAlgoLined
	\KwResult{Refined mesh}
	refinement = True\;
	\While{refinement == True}{
		refinement =False\;
		\For{each element in mesh}{
			\If{$\epsilon_e < \epsilon_\text{refine}$}{
				\If{$h > f({\epsilon})$ or $h > h_\text{min}$}{
					refine cell\;
					refinement =True\;
				}
				
			}
		}
	}
	\caption{Mesh refinement}
\end{algorithm}

\begin{wrapfigure}{r}{0.4\textwidth}
	\centering
	\def\svgwidth{0.3\columnwidth}
	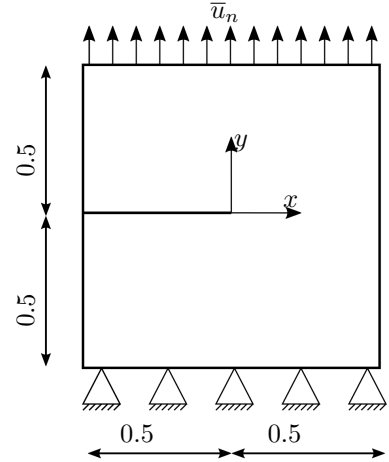

	\caption{Benchmark problem: a linear elastic panel with an edge crack  under mode I loading. All units are mm.}
	\label{fig:2d problem}
\end{wrapfigure}
Care needs to be taken to properly transfer the boundary conditions from the coarse to the fine mesh. Additionally, to ensure there are no artefacts, it is necessary to guarantee a good quality of the refined mesh, i.\,e., that aspect ratio, interior angles and mesh topology are appropriate. This refinement algorithm is called recursively until no further refinement is required. Henceforth, the spatially varying length variable $\epsilon(\bx)$ is used in conjunction with this adaptive mesh refinement strategy.

\subsection{Numerical study}\label{sec:numerical implementation}
We analyse convergence with the aid of the standard benchmark problem of an edge crack in a linear elastic panel, loaded in mode I through by applying  uniform displacements $\overline{u}_n$ at the top boundary. A schematic sketch of the problem is shown in Figure~\ref{fig:2d problem}.\\

To solve the coupled field problem, we follow the history-based strategy of \cite{miehePhaseFieldModel10}, now extended with the computation of $\epsilon(\bx)$.
All the individual steps are briefly explained below and combined in Algorithm \ref{alg: staggered algorithm}.
Here, $n$ denotes the actual load step and $M$ the total number of load steps.

1. Initialization:  The displacement field $\boldsymbol{u}_{n-1}$, phase-field $c_{n-1}$, regularisation length $\epsilon_{n-1}$ and history field $\mathcal{H}_{n-1}$ at load step ${n-1}$ are given. Update prescribed load $\overline{{u}}_{n}$ at load step ${n}$.

2. Solve for displacement field: Determine the new displacement $\boldsymbol{u}_{n}$ from equation \eqref{eq:stress} and boundary conditions with given phase-field $c_{n-1}$ and regularisation length $\epsilon_{n-1}$.

3. Compute history: Determine the maximum strain energy obtained in history
$$
\mathcal{H}_{n}=\left\{\begin{array}{ll}
	\psi\left( \boldsymbol{\varepsilon}_{n}\right) & \text { for } \psi\left( \boldsymbol{\varepsilon}_{n}\right)>\mathcal{H}_{n} \\
	\mathcal{H}_{n-1} & \text { otherwise}
\end{array}\right.
$$
and store it as a history field over the domain $\Omega$.

4. Solve for phase field: Determine the current fracture phase-field $c_{n}$ from equation \eqref{eq:phasefield} and boundary conditions with given displacement $u_n$ and regularisation length $\epsilon_{n-1}$.

5. Determine $\epsilon$: Compute the current $\epsilon_{n}$ from equation \eqref{eq:spatial epsilon}.\\

\begin{algorithm}[H]
	\label{alg: staggered algorithm}
	\SetAlgoLined
	\KwResult{Solution $\boldsymbol{u}$, $c$ and $\epsilon$}
	initialization\;
	\While{n $<$ M}{
		\While{error $>$ tolerance}{
			solve for $\boldsymbol{u}_n$\;
			solve for $c_n$\;
			compute $\epsilon_n$\;
		}
		\If{adaptivity == True}{
			$ refinement $ = refine mesh\;
		}
		\If{refinement == False}{
			$\overline{u}_{n+1} = \overline{u}_{n} + \Delta\overline{u}$\;
			n=n+1\;
		}
	}
	\caption{Solution procedure.}
\end{algorithm}

The described staggered approach is adopted to solve the problem numerically. The $adaptivity$ parameter is set to $True$ only for the spatially varying length variable scenario.
In addition to the adaptive mesh refinement strategy, we use the ``regular cut'' refinement algorithm of the FEniCS framework \cite{alnaesFEniCSProjectVersion15} to carry out the refinement.


\subsubsection{Numerical parameters}\label{sec:numerical implementation}
The material of the edge-crack panel has  the Lam\'{e} parameters $\mu = 80.7692 \cdot 10^3\,$MPa,  $\lambda = 121.1538 \cdot 10^3\,$MPa, and a fracture toughness of $\mathcal{G}_c = 2.7\  \text{MPa\,mm} $. Henceforth all parameters are understood in the units mm and MPa or combinations thereof.

We choose the following load steps to ensure the convergence of the fixed point iterations in the staggered approach.
\begin{align*}
	\Delta \overline{u} := \begin{cases}
		7 \cdot 10^{-4}  \quad \overline{u}_n  < 4.9 \cdot 10^{-3}, \\
		7 \cdot 10^{-7} \quad \text{otherwise.}
	\end{cases}
\end{align*}
In addition, the parameters for mesh refinement are $\epsilon_\text{refine}= 0.75 \epsilon_a$ and $h_\text{min} = \nicefrac{h}{8}$ where $\epsilon_a$ is the value of $\epsilon$ away from the crack tip and $h$ is the initial uniform mesh size. Apart from these, the penalty and the model parameter are estimated.

\subsubsection {Parameter estimation} \label{section:parameter estimation}
We estimate the required parameters $\beta$ and $\eta$ based on the general profile of the phase-field $c$. It is clear that, away from the crack, the specimen is intact, i.\,e., $c=0$. Furthermore, due to the boundary condition $\nabla c \cdot \bn =0$, the phase-field smoothes out towards the boundary, i.\,e.,  $  |\nabla c| $ approaches zero and the length variable can be large. Thus the choose $\epsilon_a = 10 h$ away from the crack. This is substituted in equation \eqref{eq:spatial epsilon} which leads to the following relation between $\beta$ and $\eta$
\begin{align}
	\eta = \frac{200 h^2 \beta}{\mathcal{G}_c} .
	\label{eq:eta}
\end{align}
Similarly, at the crack tip, the specimen is damaged, i.\,e., $c=1$. Furthermore, along the crack there is the transition zone from damaged to undamaged material. Numerical experiments suggest that the magnitude of the phase-field gradient   $|\nabla c|$ in this region is of the order $ \nicefrac{1}{h}$. Thus, a smaller regularisation length that can be resolved by the mesh is needed, and we choose $\epsilon_c = 2h$. Substituting these values and equation \eqref{eq:eta} in \eqref{eq:spatial epsilon} results in
\begin{align}
	\beta = \frac{3 \mathcal{G}_c }{192 h^2}  .
\end{align}
For the specific case of $\mathcal{G}_c = 2.7$ and $h= 0.01$ it yields $\beta = 421.875$ and $\eta= 3.125$. We remark that the parameters $\epsilon_a$ and $\epsilon_c$ are only used to obtain an estimate of the value of $\eta$ and $\beta$; they are not used for the actual simulations.

\subsection{Results}\label{sec:results}
In the following section, we solve the problem for an optimal uniform length variable $\epsilon$ and a spatially varying length variable $\epsilon(\boldsymbol{x})$ in a staggered manner as defined in Algorithm~\ref{alg: staggered algorithm}. The numerical parameters for the simulations are either given or estimated in the previous sections. Few parameters like the initial mesh size $h$ and the minimum allowed mesh size $h_\text{min}$ are defined along with the different numerical experiments. To investigate how the solution of the phase-field model approximates the solution of discrete fracture, we study convergence for $h$ and $\epsilon$ approaching zero. However, we first look at the convergence behavior for decreasing $h$ for various given fixed $\epsilon$ in the following section.

\subsubsection{Convergence with respect to $h$}\label{sec:fun of epsilon}
\begin{figure}
	\begin{minipage}{.3\textwidth}
		\centering
		\begin{tabular}{|c | c|}
			\hline
			mesh size $h$ & $\epsilon$\\
			\hline
			0.024 & 0.1\\
			0.012  & 0.0875\\
			0.006 & 0.075\\
			0.003 & 0.0625\\
			0.0015 & 0.05\\
			\hline
		\end{tabular}
		\captionsetup{type=table}
		\caption{Combinations of $h$ and $\epsilon$ for the numerical study.}
		\label{tab:h and eps combination}
	\end{minipage}%
	\begin{minipage}{0.69\textwidth}
		\centering
		\fontsize{9}{7} \selectfont
	\def\svgwidth{1\columnwidth}
	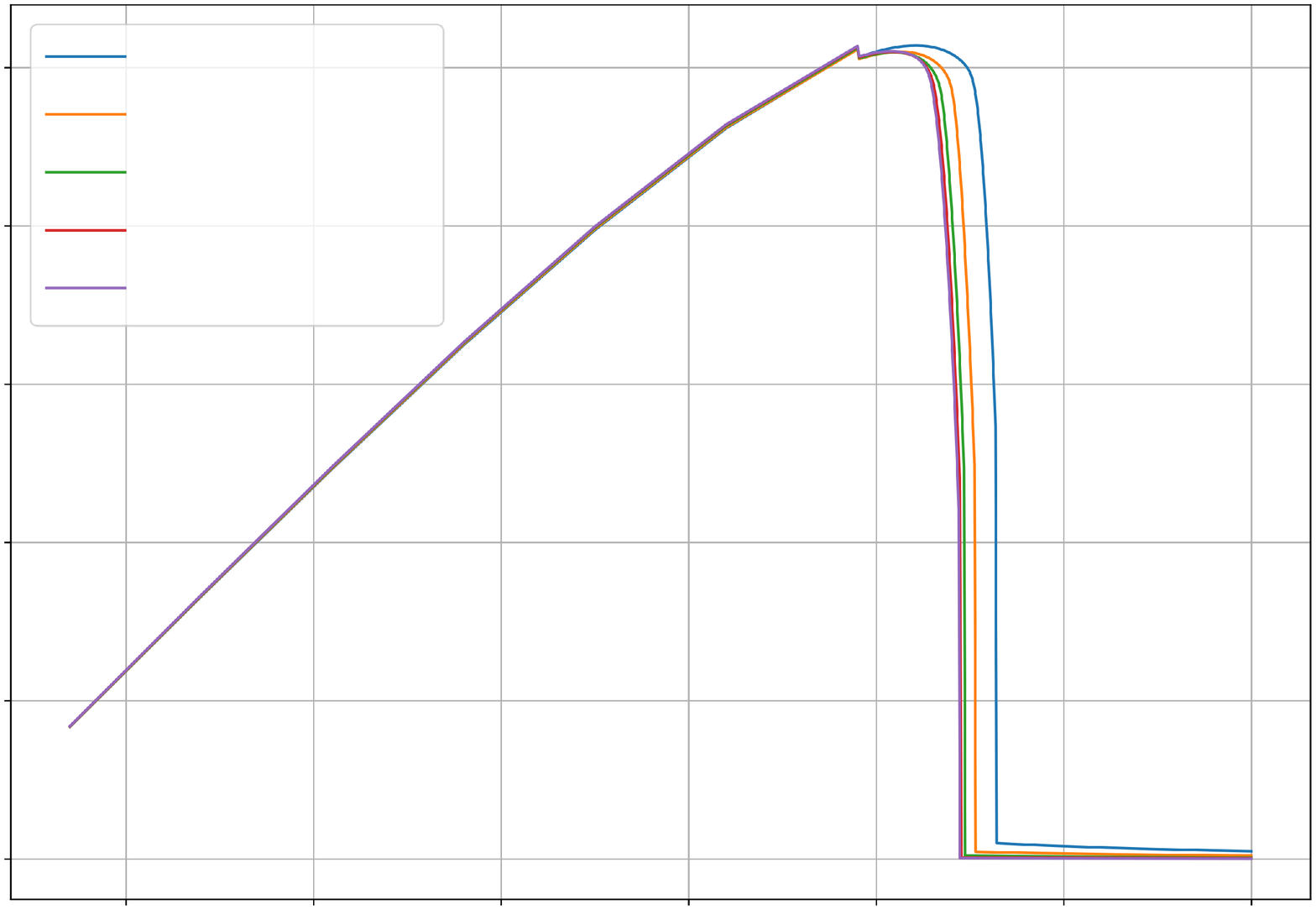

		\caption{Force {\sl vs.}~displacement in the edge-crack panel for $\epsilon = 0.1$.}
		\label{fig: force vs disp eps 0.1}
	\end{minipage}
\end{figure}

In  \cite{bourdinVariationalApproachFracture08a} Bourdin suggest a  co-relation between $\epsilon$, $h$, $\mathcal{G}_c$ and $\mathcal{G}_\text{eff}$ as $\mathcal{G}_{\text{eff}} = {\mathcal{G}_c}((1+\nicefrac{h}{4 \epsilon})$. Here, $\mathcal{G}_{\text{eff}}$ represents the overestimation of the material's  fracture toughness $\mathcal{G}_c$ in the phase-field model. This relation holds  for the specific case of 2D linear finite elements, and in \cite{eggerDiscretePhaseField19,phamExperimentalValidationPhasefield17} realistic results are obtained using this corrected value $\mathcal{G}_\text{eff}$. However, this does not always lead to a converged force versus displacement curve for a decreasing mesh size $h$.
Hence, we investigate the force-displacement relation for decreasing $h$ and a constant given $\epsilon$ for all possible combinations of $h$ and $\epsilon$ listed in Table \ref{tab:h and eps combination}. 
The computations show that a relatively fine mesh is required for a  converged result. For illustration, the force versus displacement curve for $\epsilon =0.1$  is displayed in Figure \ref{fig: force vs disp eps 0.1}. It suggests that the mesh size $h$ should be at the least $\nicefrac{\epsilon}{17}$ or smaller.
This observation is then used for mesh refinement, i.\,e., $f(\epsilon) = \nicefrac{\epsilon}{17}$.

\subsubsection{Convergence with respect to $h$ and $\epsilon$}

\begin{figure}[h!]
	\begin{center}

		\fontsize{9}{8} \selectfont
	\def\svgwidth{0.69\columnwidth}
	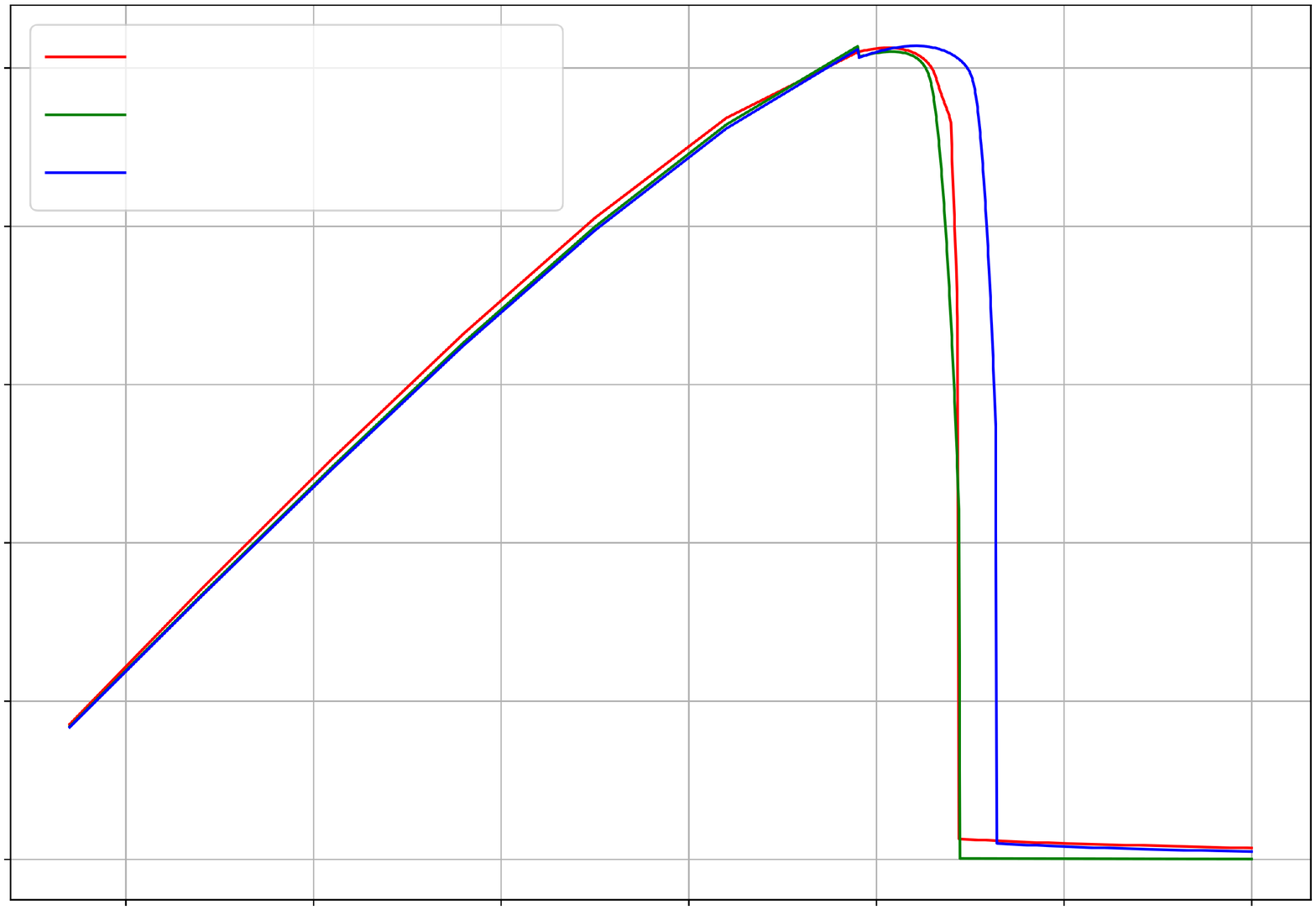

		\caption{Force {\sl vs.}~displacement in the edge-crack panel for spatially adaptive and standard phase-field model solutions. }
		\label{fig:force vs disp spatial eps}
	\end{center}
\end{figure}

\begin{figure}[h!]
	\begin{subfigure}{0.45\textwidth}
		\includegraphics[width=\textwidth]{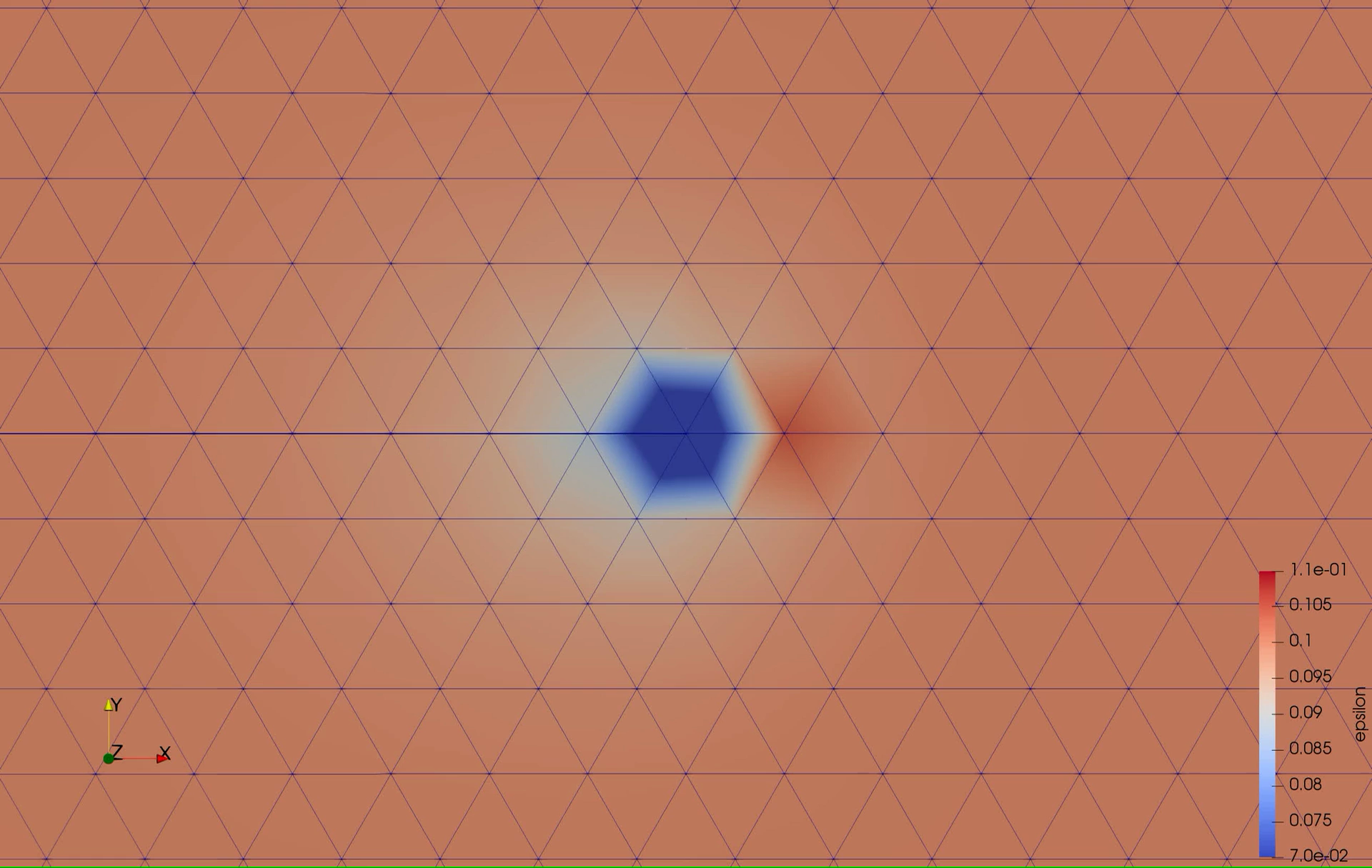}
	\end{subfigure}
	\begin{subfigure}{0.45\textwidth}
		\includegraphics[width=\textwidth]{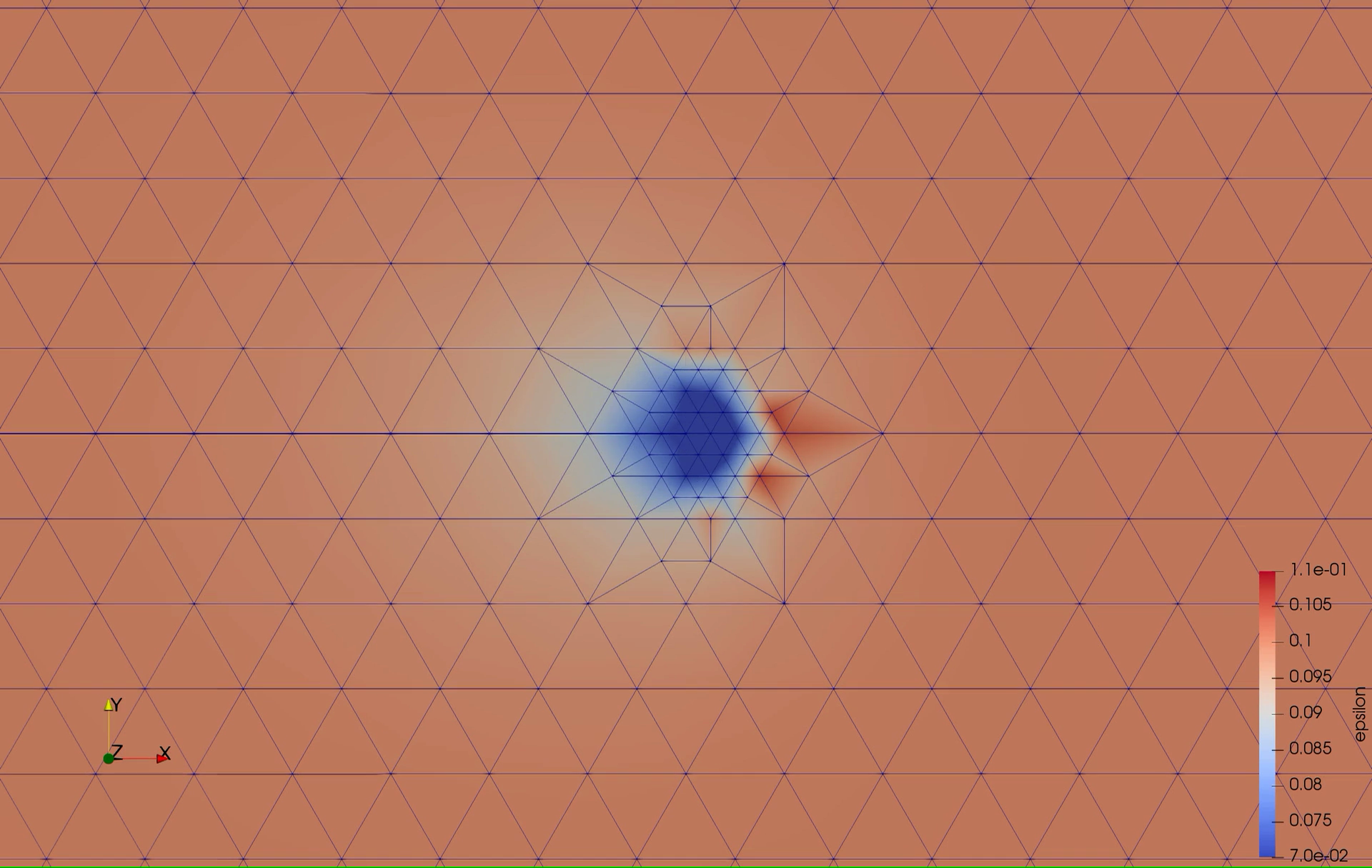}
	\end{subfigure}
	\vspace{2ex}
	%
	\begin{center}
		\includegraphics[width=0.5\textwidth]{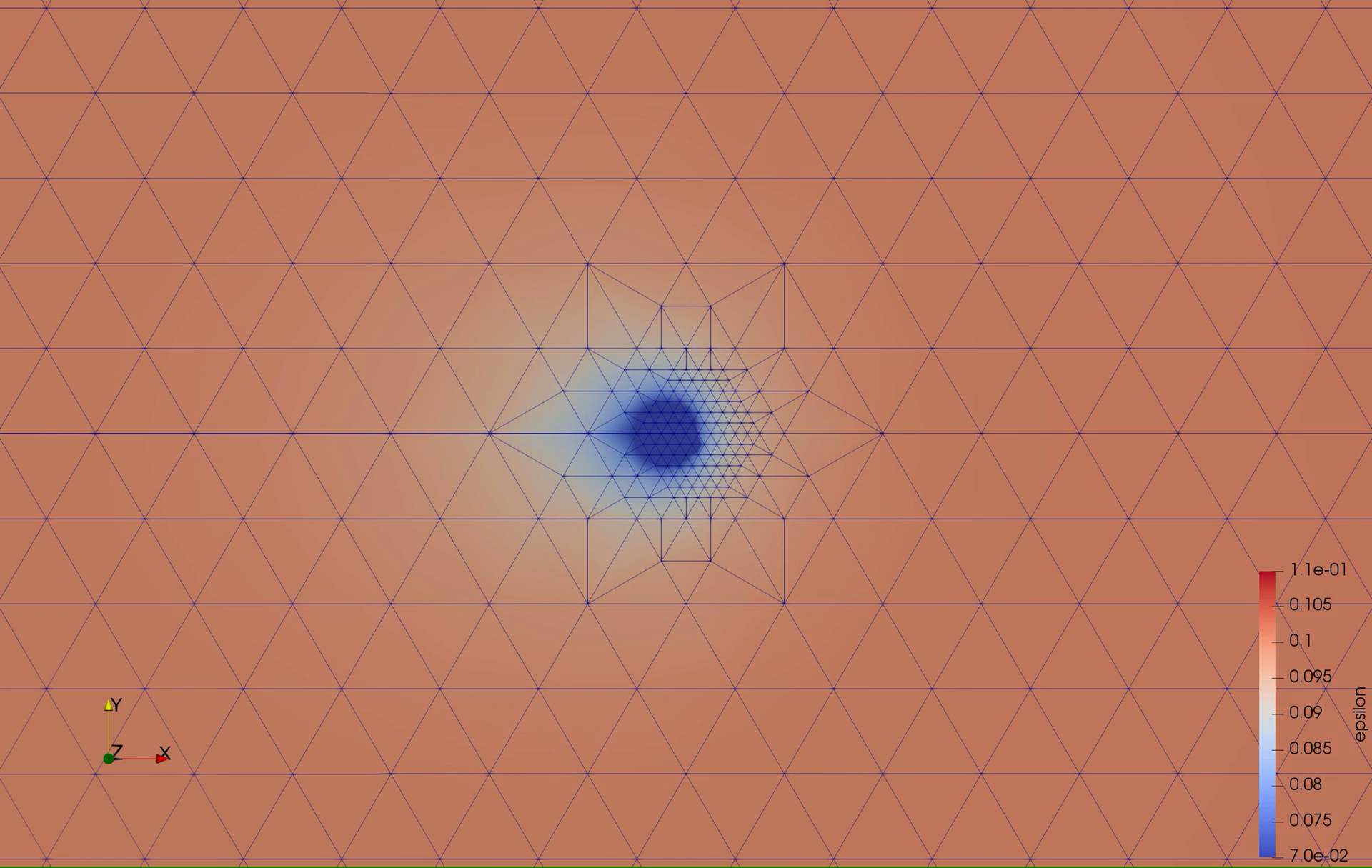}
	\end{center}
	\caption{Mesh refinement at $\overline{u}_n=0.0014$ and contour plot for the length variable $\epsilon(\bx)$ around the crack tip of the edge-crack panel; blue and red regions represent small and large values of $\epsilon$. }
	\label{fig:mesh refine}
\end{figure}

\begin{figure}[h!]
	\begin{subfigure}{0.32\textwidth}
		\includegraphics[width=\textwidth]{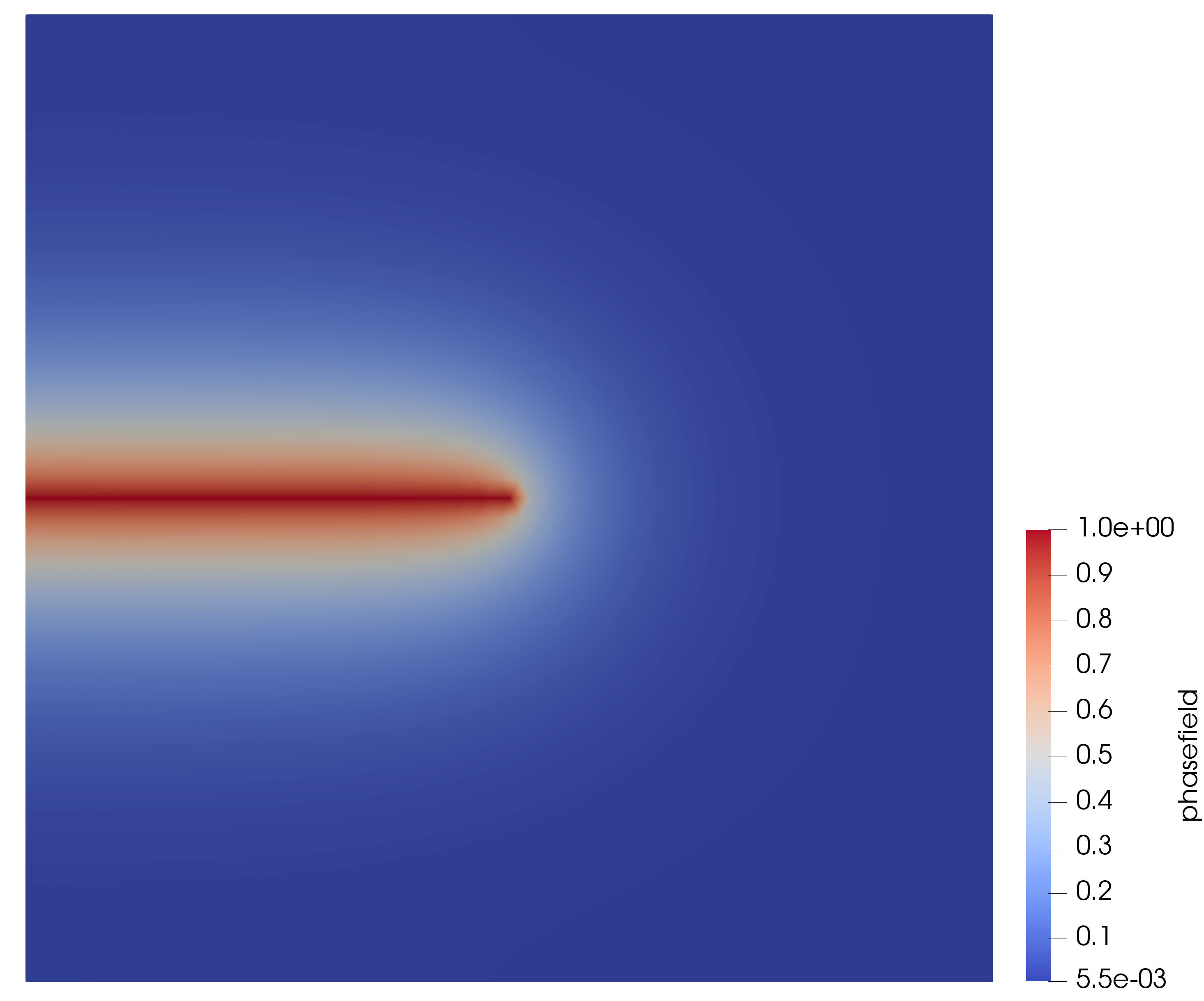}
	\end{subfigure}
	\begin{subfigure}{0.32\textwidth}
		\includegraphics[width=\textwidth]{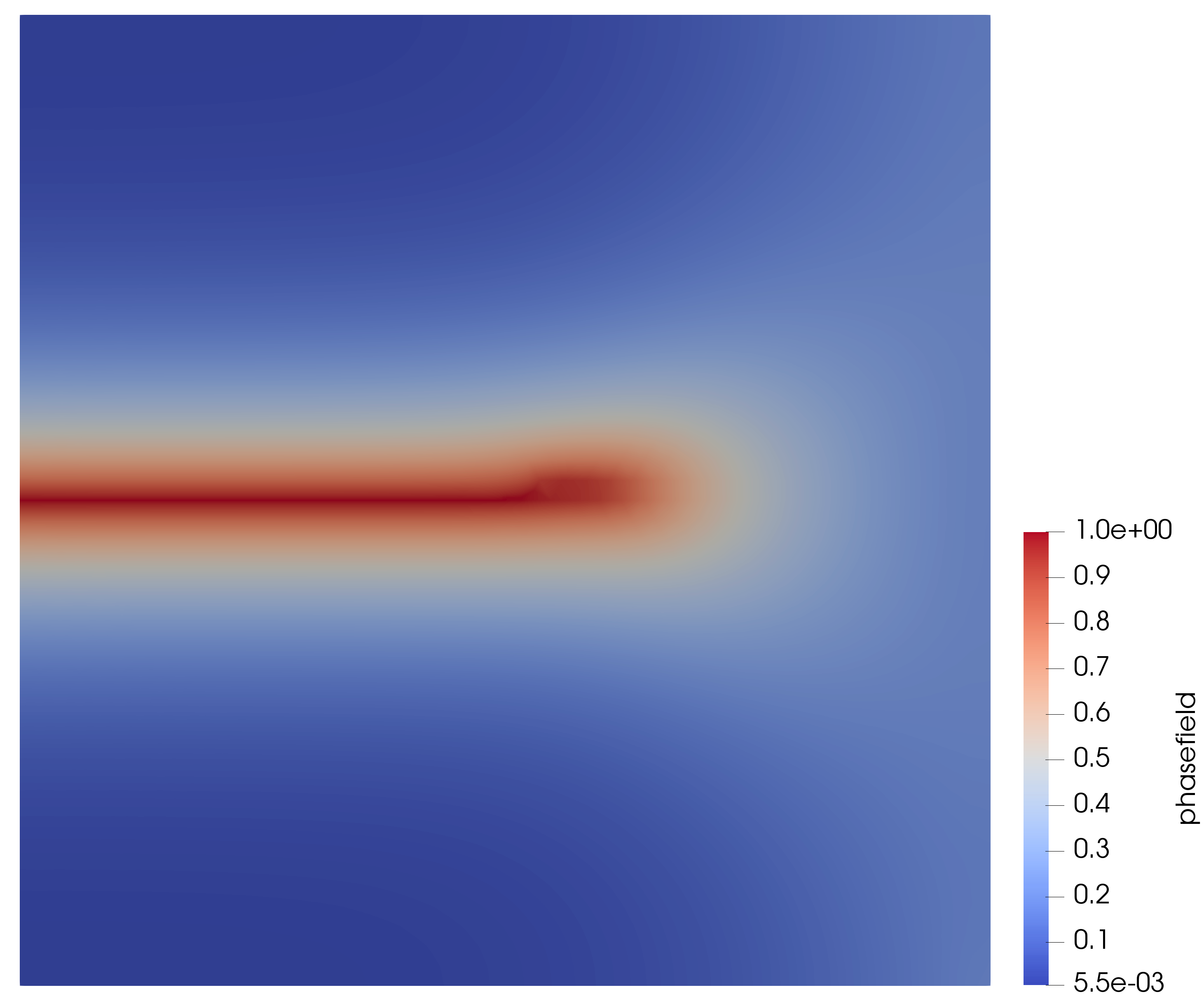}
	\end{subfigure}
	\begin{subfigure}{0.32\textwidth}
		\includegraphics[width=\textwidth]{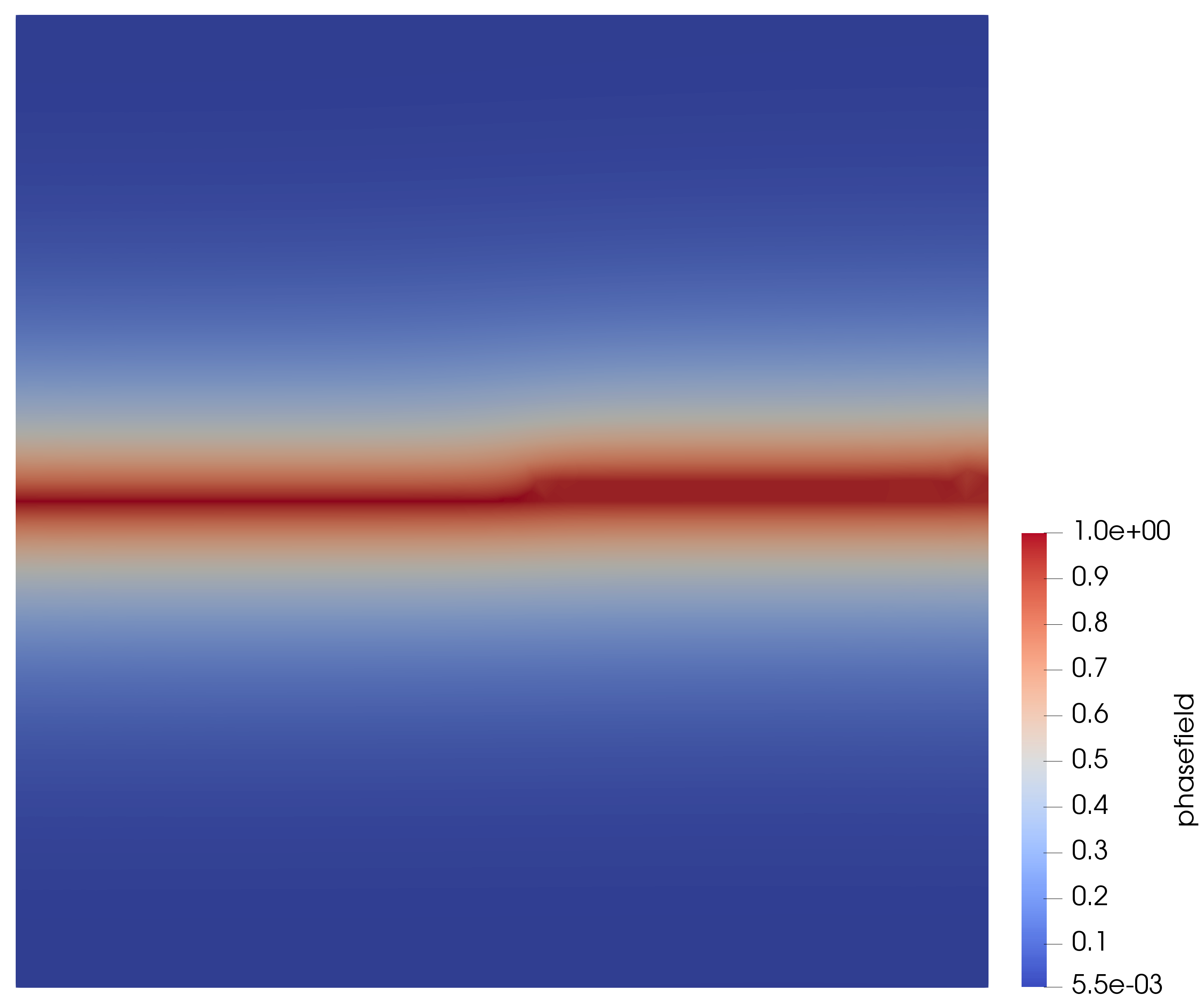}
	\end{subfigure}
	\vspace{2ex}
	\caption{Evolution of the phase-field in the  edge-crack panel at load steps $\overline{u}_n = 0.0007$, $\overline{u}_n = 0.005425$ and $\overline{u}_n = 0.005439$; the red color represents the damaged region with $c\geq 0.9$. }
	\label{fig:phase-field}
\end{figure}

We compare the resulting force-displacement curves for the spatially varying length variable $\epsilon(\boldsymbol{x})$ to that of a given constant $\epsilon = 0.1$. The spatially varying length variable $\epsilon(\boldsymbol{x})$ is solved with an initial mesh of $h = 0.024$, $h_\text{min} = 0.003$, $\beta = 540$, $\eta = 4$ and the other parameters as defined before.
We start with a coarse mesh for $\epsilon(\boldsymbol{x})$. Figure \ref{fig:force vs disp spatial eps} shows that the curves are comparable to that of constant $\epsilon =0.1$ of a fine mesh $h=0.0015$.
This result already indicates a significant improvement   without much increase in computational costs, i.\,e., adaptive mesh refinement occurs only at the crack tip. The mesh refinement at load step $u_0 = 0.0014$ can be seen in Figure \ref{fig:mesh refine}. The corresponding evolution of the phase-field is depicted in Figure \ref{fig:phase-field}. However, this is a specific case which is shown here exemplarily.

\begin{figure}[h!]
	\begin{center}
		\fontsize{9}{7} \selectfont
	\def\svgwidth{1.\columnwidth}
	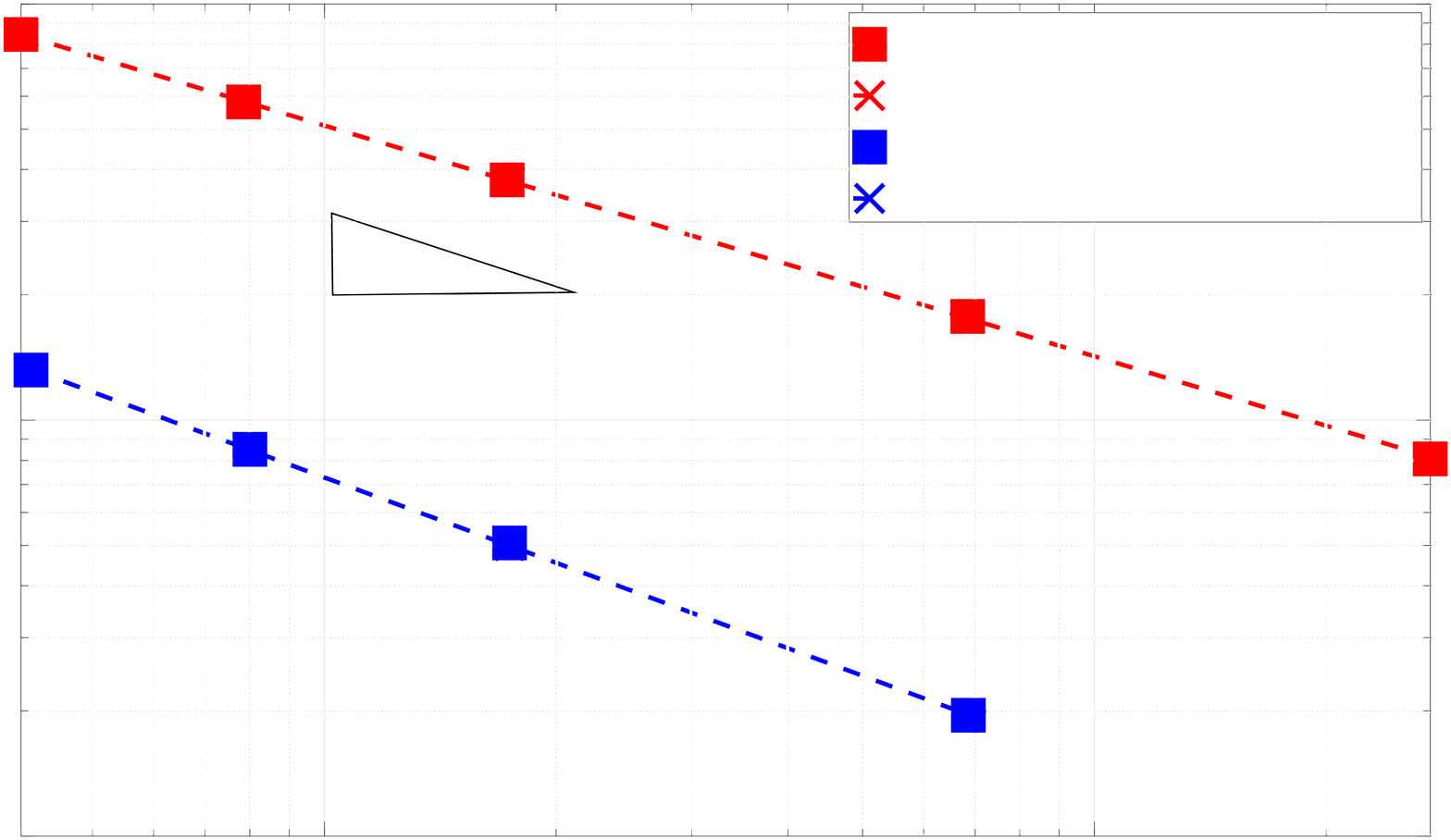

		\caption{Energy convergence 
			for the spatially varying length variable and the optimal uniform length variable solutions. }
		\label{fig:comparison uniform eps and spatial eps}
	\end{center}
\end{figure}

Hence, to show convergence of the solution, we evaluate the error in total energy for an increasing number of degrees of freedom. For a fair comparison, we compare the results of the spatially varying length variable $\epsilon(\boldsymbol{x})$ with that of the optimal uniform length variable $\epsilon$. These two problems are solved for initially uniform mesh sizes  $ h \in \{0.0015, 0.003, 0.006, 0.009, 0.0125 \}$. The parameters $\beta=2160, \ \eta=4$ and $h_\text{min} = h/8$ are used.
We validate the results in the linear regime for a particular load step of prescribed displacement of $\overline{u}_n = 0.0042$. The optimal uniform length variable $\epsilon$ and the spatially varying length variable $\epsilon(\boldsymbol{x})$ in both cases is proportional to $1/|\nabla c|$. Hence, as we decrease $h$, the value of $\epsilon$ also decreases. This ensures that we study convergence to a discrete crack solution as both $h$ and $\epsilon$ tend to zero.

For a convergence plot we fit the values of the total energy $E$ at $\overline{u}_n=0.0042$ and the total number of degrees of freedom $N$ with a power law, $E - E_\text{min} = C N^a$, where $E_\text{min}, \ C\ \text{and}\ a$ are fitting parameters. Here, $a$ represents the convergence rate and $E_\text{min}$ is the minimum energy in the limit as $h$ approaches 0.
Our fitting results   in $E_\text{min} = 218.48$, $C = 3$ and  $ a =-0.554$ for the optimal uniform length variable $\epsilon$ and $E_\text{min} = 218.44$, $ C = 1.7$ and $ a =-0.657$ for the spatially adaptive $\epsilon(\boldsymbol{x})$. Hence, it is evident that we obtain a smaller constant and a slightly steeper decrease of the error in the case of a spatially varying length variable $\epsilon(\boldsymbol{x})$ as expected from the adaptive mesh refinement strategy. The curves in Figure \ref{fig:comparison uniform eps and spatial eps} shows  relatively large reduction in error $E - E_\text{min}$  for any given particular mesh size.

\section{Summary and concluding remarks}\label{sec:summary}
We have developed a phase-field model of fracture that sets forth a variational basis for spatial adaptivity. The model generalizes the conventional phase-field model based on the Ambrosio and Tortorelli functional \cite{ambrosioApproximationFunctionalDepending90,giacominiAmbrosioTortorelliApproximationQuasistatic05} by allowing a spatially varying length variable $\epsilon (\bx)$ in the energy functional. The optimal regularisation length is then obtained by energy minimization in the same manner as the displacement and phase fields. We utilise the extended phase-field model as a basis for a mesh refinement strategy, whereby the mesh size is required to resolve the optimal regularisation length locally. We implement the resulting solution procedure in the framework of the finite element library FEniCS. We have assessed the convergence properties of the spatially varying length variable $\epsilon(\bx)$ and optimal uniform length variable $\epsilon$ solutions by means of the standard benchmark problem of an edge-crack linear-elastic panel loaded in mode I under displacement control. The numerical tests show that the adaptive solutions exhibit the same convergence rate as the conventional phase-field model, as expected, albeit with a much reduced constant. As a result, the optimal uniform length variable $\epsilon$ solutions require much smaller meshes than the spatially varying length variable $\epsilon(\bx)$ solutions at equal accuracy.

We emphasise that, in this study, we regard phase-field solutions as approximations of Griffith fracture. In particular, we expect convergence, as the regularisation length $\epsilon$ tends to zero, to mathematically sharp crack solutions for static cracks and to Griffith solutions, governed by a specific fracture energy $\mathcal{G}_c$, for growing cracks. We, therefore, understand convergence specifically as convergence to the Griffith solution as {\sl both} $\epsilon$ and $h$ tend to zero. For uniform $\epsilon$, the optimal way of approaching that joint limit has been elucidated analytically by \cite{bellettiniDiscreteApproximationFree94} and numerically by \cite{pandolfiComparativeAccuracyConvergence21}. The present work extends this strategy to the spatially varying length variable case and, in particular, determines the entire optimal spatial distribution of the regularisation length for a given number of nodes.

In closing, we emphasise that a limited and selected set of numerical tests has been used in the present study. Additional tests, including tests in three dimensions, are greatly to be desired. We also point out that mesh adaptivity, despite extensive and sustained work, remains challenging and is not a standard feature in many commercial finite element codes. An implementation of spatially adaptive phase-field models based on meshless methods suggests itself as a natural direction for further research.

\textbf{Acknowledgement}: The German Research Foundation (DFG) is gratefully acknowledged for funding this research within the research training group GRK2423 FRASCAL.\\
KW and MO gratefully acknowledge the support of the DFG within the Priority Program~2256 ``Variational Methods for Predicting Complex Phenomena in Engineering Structures and Materials'', project number 422730790.

\bibliography{spatially_adaptive_phasefield_model}

\end{document}

%% file: package.tex
\usepackage{fullpage}

\usepackage[parfill]{parskip}

\usepackage{amsfonts,amsmath,amssymb}

\usepackage[nomargin,inline,marginclue,draft]{fixme}
\fxsetup{theme=color}
\definecolor{fxtarget}{rgb}{0.8000,0.0000,0.0000}

\usepackage{graphicx}

\usepackage{grffile}

\usepackage{import}
\usepackage{epstopdf}

\usepackage{subcaption}

\usepackage{wrapfig}

\usepackage[ruled,vlined]{algorithm2e}

\usepackage{nicefrac}

\newcommand{\bx}{\boldsymbol{x}}
\newcommand{\bu}{\boldsymbol{u}}
\newcommand{\bn}{\boldsymbol{n}}
\newcommand{\bvarepsilon}{\boldsymbol{\varepsilon}}

\definecolor{dgreen}{RGB}{26,102,46}
\definecolor{orange}{RGB}{255,140,0}

\newcommand{\td}{\,\mathrm{d}}

\usepackage{epstopdf}

%% file: figure/2d_problem_def.eps_tex
\begingroup%
  \makeatletter%
  \providecommand\color[2][]{%
    \errmessage{(Inkscape) Color is used for the text in Inkscape, but the package 'color.sty' is not loaded}%
    \renewcommand\color[2][]{}%
  }%
  \providecommand\transparent[1]{%
    \errmessage{(Inkscape) Transparency is used (non-zero) for the text in Inkscape, but the package 'transparent.sty' is not loaded}%
    \renewcommand\transparent[1]{}%
  }%
  \providecommand\rotatebox[2]{#2}%
  \newcommand*\fsize{\dimexpr\f@size pt\relax}%
  \newcommand*\lineheight[1]{\fontsize{\fsize}{#1\fsize}\selectfont}%
  \ifx\svgwidth\undefined%
    \setlength{\unitlength}{227.1207273bp}%
    \ifx\svgscale\undefined%
      \relax%
    \else%
      \setlength{\unitlength}{\unitlength * \real{\svgscale}}%
    \fi%
  \else%
    \setlength{\unitlength}{\svgwidth}%
  \fi%
  \global\let\svgwidth\undefined%
  \global\let\svgscale\undefined%
  \makeatother%
  \begin{picture}(1,1.23552363)%
    \lineheight{1}%
    \setlength\tabcolsep{0pt}%
    \put(0,0){\includegraphics[width=\unitlength]{2d_problem_def.eps}}%
    \put(0.52780218,1.18200034){\color[rgb]{0,0,0}\makebox(0,0)[lt]{\lineheight{1.25}\smash{\begin{tabular}[t]{l}$\overline{u}_n$\end{tabular}}}}%
    \put(0.68142218,0.04267874){\color[rgb]{0,0,0}\makebox(0,0)[lt]{\lineheight{1.25}\smash{\begin{tabular}[t]{l}$0.5$\end{tabular}}}}%
    \put(0.28515678,0.04267874){\color[rgb]{0,0,0}\makebox(0,0)[lt]{\lineheight{1.25}\smash{\begin{tabular}[t]{l}$0.5$\end{tabular}}}}%
    \put(0.05321963,0.35217727){\color[rgb]{0,0,0}\rotatebox{87.516235}{\makebox(0,0)[lt]{\lineheight{1.25}\smash{\begin{tabular}[t]{l}$0.5$\end{tabular}}}}}%
    \put(0.0532196,0.75504793){\color[rgb]{0,0,0}\rotatebox{87.516235}{\makebox(0,0)[lt]{\lineheight{1.25}\smash{\begin{tabular}[t]{l}$0.5$\end{tabular}}}}}%
    \put(0.72263243,0.67593686){\color[rgb]{0,0,0}\makebox(0,0)[lt]{\lineheight{1.25}\smash{\begin{tabular}[t]{l}$x$\end{tabular}}}}%
    \put(0.59136962,0.83939612){\color[rgb]{0,0,0}\makebox(0,0)[lt]{\lineheight{1.25}\smash{\begin{tabular}[t]{l}$y$\end{tabular}}}}%
  \end{picture}%
\endgroup%

%% file: figure/epsilon_0.1.eps_tex
\begingroup%
  \makeatletter%
  \providecommand\color[2][]{%
    \errmessage{(Inkscape) Color is used for the text in Inkscape, but the package 'color.sty' is not loaded}%
    \renewcommand\color[2][]{}%
  }%
  \providecommand\transparent[1]{%
    \errmessage{(Inkscape) Transparency is used (non-zero) for the text in Inkscape, but the package 'transparent.sty' is not loaded}%
    \renewcommand\transparent[1]{}%
  }%
  \providecommand\rotatebox[2]{#2}%
  \newcommand*\fsize{\dimexpr\f@size pt\relax}%
  \newcommand*\lineheight[1]{\fontsize{\fsize}{#1\fsize}\selectfont}%
  \ifx\svgwidth\undefined%
    \setlength{\unitlength}{817.58435059bp}%
    \ifx\svgscale\undefined%
      \relax%
    \else%
      \setlength{\unitlength}{\unitlength * \real{\svgscale}}%
    \fi%
  \else%
    \setlength{\unitlength}{\svgwidth}%
  \fi%
  \global\let\svgwidth\undefined%
  \global\let\svgscale\undefined%
  \makeatother%
  \begin{picture}(1,0.72531183)%
    \lineheight{1}%
    \setlength\tabcolsep{0pt}%
    \put(0,0){\includegraphics[width=\unitlength]{{epsilon_0.1}.eps}}%
    \put(0.17840323,0.04533743){\makebox(0,0)[t]{\lineheight{1.25}\smash{\begin{tabular}[t]{c}0.001\end{tabular}}}}%
    \put(0.30643322,0.04533743){\makebox(0,0)[t]{\lineheight{1.25}\smash{\begin{tabular}[t]{c}0.002\end{tabular}}}}%
    \put(0.43446323,0.04533743){\makebox(0,0)[t]{\lineheight{1.25}\smash{\begin{tabular}[t]{c}0.003\end{tabular}}}}%
    \put(0.56249322,0.04533743){\makebox(0,0)[t]{\lineheight{1.25}\smash{\begin{tabular}[t]{c}0.004\end{tabular}}}}%
    \put(0.69052317,0.04533743){\makebox(0,0)[t]{\lineheight{1.25}\smash{\begin{tabular}[t]{c}0.005\end{tabular}}}}%
    \put(0.8185532,0.04533743){\makebox(0,0)[t]{\lineheight{1.25}\smash{\begin{tabular}[t]{c}0.006\end{tabular}}}}%
    \put(0.94658315,0.04533743){\makebox(0,0)[t]{\lineheight{1.25}\smash{\begin{tabular}[t]{c}0.007\end{tabular}}}}%
    \put(0.54328872,0.01440256){\makebox(0,0)[t]{\lineheight{1.25}\smash{\begin{tabular}[t]{c}Displacement\end{tabular}}}}%
    \put(0.09110298,0.09161858){\makebox(0,0)[rt]{\lineheight{1.25}\smash{\begin{tabular}[t]{r}0\end{tabular}}}}%
    \put(0.09110298,0.19959927){\makebox(0,0)[rt]{\lineheight{1.25}\smash{\begin{tabular}[t]{r}100\end{tabular}}}}%
    \put(0.09110298,0.30757999){\makebox(0,0)[rt]{\lineheight{1.25}\smash{\begin{tabular}[t]{r}200\end{tabular}}}}%
    \put(0.09110298,0.4155607){\makebox(0,0)[rt]{\lineheight{1.25}\smash{\begin{tabular}[t]{r}300\end{tabular}}}}%
    \put(0.09110298,0.52354142){\makebox(0,0)[rt]{\lineheight{1.25}\smash{\begin{tabular}[t]{r}400\end{tabular}}}}%
    \put(0.09110298,0.63152214){\makebox(0,0)[rt]{\lineheight{1.25}\smash{\begin{tabular}[t]{r}500\end{tabular}}}}%
    \put(0.0292527,0.37948836){\rotatebox{90}{\makebox(0,0)[t]{\lineheight{1.25}\smash{\begin{tabular}[t]{c}Force\end{tabular}}}}}%
    \put(0.19922636,0.6399672){\makebox(0,0)[lt]{\lineheight{1.25}\smash{\begin{tabular}[t]{l}h = 0.024000\end{tabular}}}}%
    \put(0.19922636,0.60047051){\makebox(0,0)[lt]{\lineheight{1.25}\smash{\begin{tabular}[t]{l}h = 0.012000\end{tabular}}}}%
    \put(0.19922636,0.56097383){\makebox(0,0)[lt]{\lineheight{1.25}\smash{\begin{tabular}[t]{l}h = 0.006000\end{tabular}}}}%
    \put(0.19922636,0.52147713){\makebox(0,0)[lt]{\lineheight{1.25}\smash{\begin{tabular}[t]{l}h = 0.003000\end{tabular}}}}%
    \put(0.19922636,0.48198045){\makebox(0,0)[lt]{\lineheight{1.25}\smash{\begin{tabular}[t]{l}h = 0.001500\end{tabular}}}}%
  \end{picture}%
\endgroup%

%% file: figure/spaceEpsFvsDispComparision.eps_tex
\begingroup%
  \makeatletter%
  \providecommand\color[2][]{%
    \errmessage{(Inkscape) Color is used for the text in Inkscape, but the package 'color.sty' is not loaded}%
    \renewcommand\color[2][]{}%
  }%
  \providecommand\transparent[1]{%
    \errmessage{(Inkscape) Transparency is used (non-zero) for the text in Inkscape, but the package 'transparent.sty' is not loaded}%
    \renewcommand\transparent[1]{}%
  }%
  \providecommand\rotatebox[2]{#2}%
  \newcommand*\fsize{\dimexpr\f@size pt\relax}%
  \newcommand*\lineheight[1]{\fontsize{\fsize}{#1\fsize}\selectfont}%
  \ifx\svgwidth\undefined%
    \setlength{\unitlength}{817.58435059bp}%
    \ifx\svgscale\undefined%
      \relax%
    \else%
      \setlength{\unitlength}{\unitlength * \real{\svgscale}}%
    \fi%
  \else%
    \setlength{\unitlength}{\svgwidth}%
  \fi%
  \global\let\svgwidth\undefined%
  \global\let\svgscale\undefined%
  \makeatother%
  \begin{picture}(1,0.6977185)%
    \lineheight{1}%
    \setlength\tabcolsep{0pt}%
    \put(0,0){\includegraphics[width=\unitlength]{spaceEpsFvsDispComparision.eps}}%
    \put(0.17840323,0.04533743){\makebox(0,0)[t]{\lineheight{1.25}\smash{\begin{tabular}[t]{c}0.001\end{tabular}}}}%
    \put(0.30643322,0.04533743){\makebox(0,0)[t]{\lineheight{1.25}\smash{\begin{tabular}[t]{c}0.002\end{tabular}}}}%
    \put(0.43446322,0.04533743){\makebox(0,0)[t]{\lineheight{1.25}\smash{\begin{tabular}[t]{c}0.003\end{tabular}}}}%
    \put(0.56249321,0.04533743){\makebox(0,0)[t]{\lineheight{1.25}\smash{\begin{tabular}[t]{c}0.004\end{tabular}}}}%
    \put(0.69052317,0.04533743){\makebox(0,0)[t]{\lineheight{1.25}\smash{\begin{tabular}[t]{c}0.005\end{tabular}}}}%
    \put(0.8185532,0.04533743){\makebox(0,0)[t]{\lineheight{1.25}\smash{\begin{tabular}[t]{c}0.006\end{tabular}}}}%
    \put(0.94658315,0.04533743){\makebox(0,0)[t]{\lineheight{1.25}\smash{\begin{tabular}[t]{c}0.007\end{tabular}}}}%
    \put(0.54328872,0.01440256){\makebox(0,0)[t]{\lineheight{1.25}\smash{\begin{tabular}[t]{c}Displacement\end{tabular}}}}%
    \put(0.09110298,0.09161854){\makebox(0,0)[rt]{\lineheight{1.25}\smash{\begin{tabular}[t]{r}0\end{tabular}}}}%
    \put(0.09110298,0.19959927){\makebox(0,0)[rt]{\lineheight{1.25}\smash{\begin{tabular}[t]{r}100\end{tabular}}}}%
    \put(0.09110298,0.30757999){\makebox(0,0)[rt]{\lineheight{1.25}\smash{\begin{tabular}[t]{r}200\end{tabular}}}}%
    \put(0.09110298,0.41556069){\makebox(0,0)[rt]{\lineheight{1.25}\smash{\begin{tabular}[t]{r}300\end{tabular}}}}%
    \put(0.09110298,0.52354142){\makebox(0,0)[rt]{\lineheight{1.25}\smash{\begin{tabular}[t]{r}400\end{tabular}}}}%
    \put(0.09110298,0.63152214){\makebox(0,0)[rt]{\lineheight{1.25}\smash{\begin{tabular}[t]{r}500\end{tabular}}}}%
    \put(0.0292527,0.37948836){\rotatebox{90}{\makebox(0,0)[t]{\lineheight{1.25}\smash{\begin{tabular}[t]{c}Force\end{tabular}}}}}%
    \put(0.19922637,0.63996299){\makebox(0,0)[lt]{\lineheight{1.25}\smash{\begin{tabular}[t]{l}spatially varying $\epsilon(\bx)$\end{tabular}}}}%
    \put(0.19922637,0.60041165){\makebox(0,0)[lt]{\lineheight{1.25}\smash{\begin{tabular}[t]{l}h = 0.0015, $\epsilon$=0.1\end{tabular}}}}%
    \put(0.19922637,0.56091496){\makebox(0,0)[lt]{\lineheight{1.25}\smash{\begin{tabular}[t]{l}h = 0.024, $\epsilon$=0.1\end{tabular}}}}%
  \end{picture}%
\endgroup%

%% file: figure/comparison_of_relative_energy.eps_tex
\begingroup%
  \makeatletter%
  \providecommand\color[2][]{%
    \errmessage{(Inkscape) Color is used for the text in Inkscape, but the package 'color.sty' is not loaded}%
    \renewcommand\color[2][]{}%
  }%
  \providecommand\transparent[1]{%
    \errmessage{(Inkscape) Transparency is used (non-zero) for the text in Inkscape, but the package 'transparent.sty' is not loaded}%
    \renewcommand\transparent[1]{}%
  }%
  \providecommand\rotatebox[2]{#2}%
  \newcommand*\fsize{\dimexpr\f@size pt\relax}%
  \newcommand*\lineheight[1]{\fontsize{\fsize}{#1\fsize}\selectfont}%
  \ifx\svgwidth\undefined%
    \setlength{\unitlength}{1260bp}%
    \ifx\svgscale\undefined%
      \relax%
    \else%
      \setlength{\unitlength}{\unitlength * \real{\svgscale}}%
    \fi%
  \else%
    \setlength{\unitlength}{\svgwidth}%
  \fi%
  \global\let\svgwidth\undefined%
  \global\let\svgscale\undefined%
  \makeatother%
  \begin{picture}(1,0.56785714)%
    \lineheight{1}%
    \setlength\tabcolsep{0pt}%
    \put(0,0){\includegraphics[width=\unitlength]{comparison_of_relative_energy.eps}}%
    \put(0.27916667,0.03392857){\makebox(0,0)[lt]{\lineheight{1.25}\smash{\begin{tabular}[t]{l}10\end{tabular}}}}%
    \put(0.30416667,0.04464286){\makebox(0,0)[lt]{\lineheight{1.25}\smash{\begin{tabular}[t]{l}5\end{tabular}}}}%
    \put(0.70238095,0.03392857){\makebox(0,0)[lt]{\lineheight{1.25}\smash{\begin{tabular}[t]{l}10\end{tabular}}}}%
    \put(0.72738095,0.04464286){\makebox(0,0)[lt]{\lineheight{1.25}\smash{\begin{tabular}[t]{l}6\end{tabular}}}}%
    \put(0.39791702,0.00541667){\makebox(0,0)[lt]{\lineheight{1.25}\smash{\begin{tabular}[t]{l}Degrees of freedom N\end{tabular}}}}%
    \put(0.0827381,0.05535714){\makebox(0,0)[lt]{\lineheight{1.25}\smash{\begin{tabular}[t]{l}10\end{tabular}}}}%
    \put(0.1077381,0.06607143){\makebox(0,0)[lt]{\lineheight{1.25}\smash{\begin{tabular}[t]{l}-4\end{tabular}}}}%
    \put(0.0827381,0.28452381){\makebox(0,0)[lt]{\lineheight{1.25}\smash{\begin{tabular}[t]{l}10\end{tabular}}}}%
    \put(0.1077381,0.2952381){\makebox(0,0)[lt]{\lineheight{1.25}\smash{\begin{tabular}[t]{l}-3\end{tabular}}}}%
    \put(0.0827381,0.51309524){\makebox(0,0)[lt]{\lineheight{1.25}\smash{\begin{tabular}[t]{l}10\end{tabular}}}}%
    \put(0.1077381,0.52380952){\makebox(0,0)[lt]{\lineheight{1.25}\smash{\begin{tabular}[t]{l}-2\end{tabular}}}}%
    \put(0.06547619,0.25357143){\rotatebox{90}{\makebox(0,0)[lt]{\lineheight{1.25}\smash{\begin{tabular}[t]{l}E - E\end{tabular}}}}}%
    \put(0.07738095,0.30654762){\rotatebox{90}{\makebox(0,0)[lt]{\lineheight{1.25}\smash{\begin{tabular}[t]{l}min\end{tabular}}}}}%
    \put(0.6077381,0.5){\makebox(0,0)[lt]{\lineheight{1.25}\smash{\begin{tabular}[t]{l}E - E\end{tabular}}}}%
    \put(0.65119048,0.49047619){\makebox(0,0)[lt]{\lineheight{1.25}\smash{\begin{tabular}[t]{l}min\end{tabular}}}}%
    \put(0.67738095,0.5){\makebox(0,0)[lt]{\lineheight{1.25}\smash{\begin{tabular}[t]{l} uniform eps : a = -0.555\end{tabular}}}}%
    \put(0.60744048,0.46770833){\makebox(0,0)[lt]{\lineheight{1.25}\smash{\begin{tabular}[t]{l}power law  uniform eps\end{tabular}}}}%
    \put(0.6077381,0.44345238){\makebox(0,0)[lt]{\lineheight{1.25}\smash{\begin{tabular}[t]{l}E - E\end{tabular}}}}%
    \put(0.65119048,0.43392857){\makebox(0,0)[lt]{\lineheight{1.25}\smash{\begin{tabular}[t]{l}min\end{tabular}}}}%
    \put(0.67738095,0.44345238){\makebox(0,0)[lt]{\lineheight{1.25}\smash{\begin{tabular}[t]{l} adaptive eps: a = -0.682\end{tabular}}}}%
    \put(0.60744048,0.41116071){\makebox(0,0)[lt]{\lineheight{1.25}\smash{\begin{tabular}[t]{l}power law  adaptive eps\end{tabular}}}}%
    \put(0.35201848,0.34467052){\color[rgb]{0,0,0}\makebox(0,0)[lt]{\lineheight{1.25}\smash{\begin{tabular}[t]{l}1\end{tabular}}}}%
    \put(0.27522138,0.3859704){\color[rgb]{0,0,0}\makebox(0,0)[lt]{\lineheight{1.25}\smash{\begin{tabular}[t]{l}$a$\end{tabular}}}}%
  \end{picture}%
\endgroup%